\documentclass[pre,notitlepage,nofootinbib]{revtex4-1}
\usepackage{amsmath}
\usepackage{graphicx}

\begin{document}
\title{Fluctuations of Entropy Production in Partially Masked Electric Circuits: Theoretical Analysis}

\author{K.-H. Chiang}
\author{C.-W. Chou}
\author{C.-L. Lee*}
\author{P.-Y. Lai}
\author{Y.-F. Chen*}
\affiliation{Department of Physics, National Central University, Chungli, Taiwan}

\begin{abstract}
In this work we perform theoretical analysis about a coupled RC circuit with constant driven currents.  Starting from stochastic differential equations, where voltages are subject to thermal noises, we derive time-correlation functions, steady-state distributions and transition probabilities of the system.  The validity of the fluctuation theorem (FT)\cite{Crooks99,Seifert12} is examined for scenarios with complete and incomplete descriptions.
\end{abstract}
\maketitle

\section{Theoretical study: the complete description}
\label{sec_thy_complete}
\begin{figure}[h]
  \includegraphics[height = 4cm]{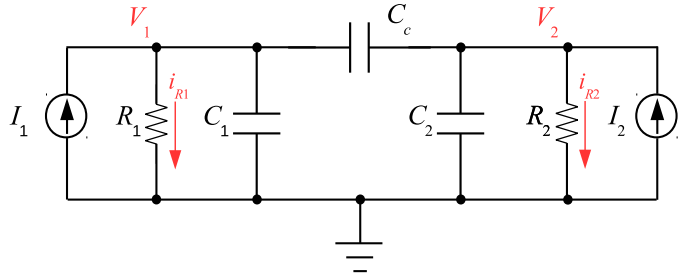}
  \caption{Illustration of the coupled RC circuit.}
  \label{Fig1}
\end{figure}

We consider the coupled RC circuit as shown in Fig.~\ref{Fig1}.  Two RC circuits of resistances and capacitances $(R_1, C_1)$ and $(R_2, C_2)$, respectively, are coupled through a third capacitance $C_c$.  The two RC circuits are subject to constant driven currents $I_1$ and $I_2$, and the voltage differences across the resistors are denoted as $V_1$ and $V_2$, respectively. 
The equation of state of this circuit is
\begin{equation}
  \hat{\bf{M}} \dot{\vec{V}} + \vec{V} - \vec{\xi} = \vec{V_d}\, ,
\end{equation}
where
$\displaystyle
\vec{V} \equiv \left( \begin{array}{c} V_1 \\ V_2 \end{array} \right),
\vec{\xi} \equiv \left( \begin{array}{c} \xi_1 \\ \xi_2 \end{array} \right),
\vec{V_d} \equiv \left( \begin{array}{c} I_1 R_1\\ I_2 R_2\end{array} \right)$, and
%\begin{equation*}
$\displaystyle \hat{\bf{M}} \equiv \left( \begin{array}{cc} R_1 (C_1+C_c) & - R_1 C_c \\ -R_2 C_c & R_2 (C_2+C_c) \end{array} \right).$
%\end{equation*}
The two resistors are thermalized at temperature $T$, while voltages across the resistors fluctuate due to the Johnson-Nyquist (thermal) noises $\xi_1$ and $\xi_2$.  The noises are assumed to be uncorrelated and Gaussian white, and they satisfy the fluctuation-dissipation relation
\begin{equation}
  \langle \vec{\xi}(s) \vec{\xi}^{\ T}(s') \rangle = \hat{\bf{\Gamma}}\delta (s-s')\, , \ \hat{\bf{\Gamma}}\equiv
  \left(\begin{array}{cc} \Gamma_1 & 0 \\ 0 & \Gamma_2 \end{array}\right)\, ,
  \label{thm_FD}
\end{equation}
with $\Gamma_m \equiv 2R_m k_B T$ for $m=1, 2$, where $k_B$ is Boltzmann's constant.

Via the change of variables $\vec{V}' \equiv \vec{V} - \vec{V_d}$ , the equation of state can be rewritten as
%\begin{equation}
$  \hat{\bf{M}} \dot{\vec{V}}' + \vec{V}' = \vec{\xi}\, ,$
%  \label{eqn_state}
%\end{equation}
which is mathematically identical to that of a non-driven circuit.  The solution of the stochastic equation is
\begin{equation}
  \vec{V}'(t) = e^{-\hat{\bf{M}}^{-1} t} \vec{V}'(0) + \int_{s=0}^{t} \! ds \, \hat{\bf{M}}^{-1} e^{-\hat{\bf{M}}^{-1} (t-s)} \vec{\xi}(s)\, .
  \label{V_sol}
\end{equation}
In our work we only focus on the steady-state condition, where the first term in Eq.~\ref{V_sol} damps out.

Using Eq.~\ref{thm_FD} and the fact that $\hat{\bf{\Gamma}}^{-1} \hat{\bf{M}}$ is symmetric, time-correlation functions between voltage signals can be derived as
\begin{align}
  & \langle \vec{V}'(t') \vec{V}'^{T}(t+t') \rangle \nonumber \\
  & = \int^{t'}_0 \!\! ds \, \hat{\bf{M}}^{-1} e^{-\hat{\bf{M}}^{-1} (t'-s)}
  \hat{\bf{\Gamma}} e^{-\hat{\bf{M}}^{-1,T} (t+t'-s)} \hat{\bf{M}}^{-1,T}\nonumber\\
  & = \int^{\infty}_0 \!\! ds \, \hat{\bf{M}}^{-2} e^{-\hat{\bf{M}}^{-1} (2t'-2s+t)}\hat{\bf{\Gamma}} \underset{t' \to \infty}{\longrightarrow} \frac12 \hat{\bf{M}}^{-1} e^{-\hat{\bf{M}}^{-1} t} \hat{\bf{\Gamma}}\, ,
  \label{V_corr}
\end{align}
and variances and covariances are just their special cases when $t \to 0$.  Using the method of diagonalization, we then derive
\begin{align}
  & \langle V'_1(t') V'_1(t'+t)\rangle = \frac{  A_1 e_1 / \lambda_1 + A_2 e_2 / \lambda_2  } {2 (\lambda_2-\lambda_1)^2} \nonumber\\
  & \langle V'_1(t') V'_2(t'+t)\rangle = \langle V'_2(t') V'_1(t'+t)\rangle = 
   \frac{  A_1 e_1 (\lambda_1-M_{11})/ \lambda_1 + A_2 e_2 (\lambda_2-M_{11})/ \lambda_2  } {2 M_{12} (\lambda_2-\lambda_1)^2} \nonumber\\
  & \langle V'_2(t') V'_2(t'+t)\rangle = \frac{  A_1 e_1 (\lambda_1 - M_{11})^2/ \lambda_1 + A_2 e_2 (\lambda_2 - M_{11})^2/ \lambda_2  } {2 M_{12}^2 (\lambda_2-\lambda_1)^2} \, ,
\end{align}
where $\lambda_1$ and $\lambda_2$ are the eigenvalues of $\hat{\bf{M}}$ ($\lambda_1$ is assigned as the larger one), $e_m \equiv e^{-t/\lambda_m}$,  $A_1 \equiv \Gamma_1 (\lambda_2 - M_{11})^2+ \Gamma_2 M_{12}^2$, $A_2 \equiv \Gamma_1 (\lambda_1 - M_{11})^2 + \Gamma_2 M_{12}^2$, and $M_{mn}$ are the $(m,n)$ element of the matrix $\hat{\bf{M}}$.  Moreover, the correlations between $\vec{V}$ and $\vec{\xi}$ can be derived as $\langle \vec{V}'(t') \vec{\xi}^{\, T}(t'+t) \rangle = 0$ (no causality) and $\langle \vec{\xi}(t') \vec{V}'^{T}(t'+t) \rangle = 2 \langle \vec{V}'(t') \vec{V}'^{T}(t+t') \rangle$.

The Fokker-Planck equation of the full circuit can be shown to be
\begin{equation}
  \partial P(\vec{V}, t) / \partial t = \nabla \cdot [\hat{\bf{M}}^{-1} \vec{V}' P(\vec{V},t)] + \frac12 \nabla \cdot \hat{\bf{M}}^{-1} \hat{\bf{\Gamma}} (\hat{\bf{M}}^{-1})^{T} \nabla P(\vec{V},t)\, ,
\end{equation}
and the steady-state distribution of $\vec{V}$ is
\begin{equation}
  P_{\rm{ss}}(\vec{V}) \sim \exp(- \vec{V}'^T \hat{\bf{\Gamma}}^{-1} \hat{\bf{M}} \vec{V}')\, .
  \label{ss_allV}
\end{equation}
For convenience we use the symbol ``$\sim$'' to denote that the equality holds up to some normalizing constant that remains invariant in the time-reversal process.  Note that in the non-driven case the expression reduces to the Boltzmann factor of the stored energy in capacitors.

The transition probability of the complete description, under an infinitesimal change in time, can be derived as
\begin{equation}
  P_F(\vec{V}(t+dt)|\vec{V}(t)) \sim e^{-dt \, \vec{\xi}^{\ T}
  \hat{\bf{\Gamma}}^{-1} \vec{\xi}/2} = 
  \exp[-dt \,  (\hat{\bf{M}}\dot{\vec{V}}+\vec{V}')^{T} \hat{\bf{\Gamma}}^{-1}
  (\hat{\bf{M}}\dot{\vec{V}}+\vec{V}')/2]\, ,
  \label{P_F_all}
\end{equation}
where $\dot{\vec{V}} \equiv [\vec{V}(t+dt)-\vec{V}(t)] /dt$, and $\vec{\xi}$ represents the required noises for such transition.  In the corresponding time-reversal process,
\begin{equation}
  P_R(\vec{V}(t)|\vec{V}(t+dt)) \sim
  \exp[-dt \, (-\hat{\bf{M}}\dot{\vec{V}}+\vec{V}'')^{T} \hat{\bf{\Gamma}}^{-1} (-\hat{\bf{M}}\dot{\vec{V}}+\vec{V}'')/2]\, ,
\end{equation}
as one replaces $\dot{\vec{V}}$ with $-\dot{\vec{V}}$ and $\vec{V}'$ with $\vec{V}'' = \vec{V} +\vec{V_d}$, the latter resulting from inversion of driven currents.  Subsequently, the net dissipation can be shown to be
\begin{align}
  dS_{Q} &= k_B \ln \frac{ P_F(\vec{V}(t+dt)|\vec{V}(t)) }{ P_R(\vec{V}(t)|\vec{V}(t+dt)) }  \nonumber \\
  &= -2k_B\, dt\, \vec{V}^T \hat{\bf{\Gamma}}^{-1}\hat{\bf{M}}\dot{\vec{V}} + 2k_B\, dt\, \vec{V}^T \hat{\bf{\Gamma}}^{-1}\vec{V_d} \nonumber \\
  &= dt\, \vec{V}\cdot \vec{i}_R/T\, ,
\end{align}
where $\vec{i}_R$ represents the currents through the two resistors.  The corresponding change in Shannon entropy is
\begin{align}
  dS_{\rm{Sh}} &= -k_B \ln \frac{ P_{\rm{ss}}(\vec{V}(t+dt)) }{ P_{\rm{ss}}(\vec{V}(t)) }
   = -k_B\, dt \left[ \frac{ d\ln P_{\rm{ss}}(\vec{V}(t)) }{ dt} \right] \nonumber \\ & = 2k_B\, dt\, \vec{V}'^T \hat{\bf{\Gamma}}^{-1} \hat{\bf{M}} \dot{\vec{V}}
\end{align}
and the total entropy change $dS_{\rm{tot}} = dS_{Q} + dS_{\rm{Sh}}$ is
\begin{align}
  dS_{\rm{tot}} & = -2k_B\,dt\, \vec{V_d}^T \hat{\bf{\Gamma}}^{-1}\hat{\bf{M}}\dot{\vec{V}} + 2k_B\, dt\, \vec{V}^T \hat{\bf{\Gamma}}^{-1}\vec{V_d} \nonumber \\
  & = dt(2\vec{V}'\cdot \vec{I_d}-\vec{\xi}\cdot \vec{I_d} + \vec{V_d}\cdot \vec{I_d})/T\, .
  \label{dS_tot}
\end{align}
For our current discussions, $\vec{V_d}$ and $\vec{I_d}$ are constants, and therefore $dS_{\rm{tot}}$ is a Gaussian random variable, while $\langle dS_{\rm{tot}}\rangle = dt\, \vec{V_d}\cdot \vec{I_d}/T = dt\, (I^2_1 R_1+ I^2_2 R_2)/T$.  One can show that
\begin{align}
  \langle dS^2_{\rm{tot}}\rangle - \langle dS_{\rm{tot}} \rangle^2 & =
  (dt)^2 \langle (2\vec{V}'\cdot \vec{I_d} - \vec{\xi}\cdot \vec{I_d})^2\rangle /T^2
  \nonumber \\ &= 2k_B\, dt \vec{V_d}\cdot\vec{I_d}/T\, = 2k_B\, \langle dS_{\rm{tot}}\rangle\, .
\end{align}
It is straightforward
\footnote{Consider a Gaussian random variable $x$ of average $x_0$ and width $\sigma_x$.  The corresponding symmetry function is $\mathrm{Sym}(x) = \ln [ P(x) / P(-x)] = 2x_0 x/ \sigma^2_x $.  Thus Gaussianity is a sufficient condition of the FT-like behavior.  FT is valid if $\sigma^2_x = 2x_0$ (in dimensionless units; in cases where $x$ means entropy then FT validates if $\sigma^2_x = 2k_B x_0$), and for the case where the observed slope is not equal to 1, FT can be easily restored with the rescaled variable $x' = (2x_0/\sigma^2_x) x$.}
to demonstrate that FT holds for any Gaussian random variable whose ratio of variance over mean value is $2k_B$
.  Moreover, with the aid of time-correlation functions, one can also demonstrate the validity of FT over finite-time processes, where $\Delta S_{\rm{tot},\tau} = \int^{\tau}_{t=0} dS_{\rm{tot}}$, and $\langle \Delta S_{\rm{tot},\tau}\rangle = \tau\, \vec{V_d}\cdot \vec{I_d}/T$.

\section{Reduced description (A): naive description}
\label{sec_apparent_S}
In the reduced descriptions, we neglect the signal $V_2$ intentionally, and we would like to check whether FT can still validate with the knowledge of $V_1$ only.  Note that since the current throught $R_1$ is not measured, the actual dissipation through the resistor is not known, while there exist many methods towards guessing an effective dissipation simply from the time series of $V_1$.  In this work we adopt two methods.  In description (A), we treat the time series of $V_1$ as that from a virtual single-RC circuit, and compute the current and therefore dissipation directly following the equation of this simplified circuit.  And in description (B), an effective dissipation can be derived using the ratio of forward and backward transition probabilities in $V_1$ over infinitesimal timesteps. 

We first derive the steady-state probability distribution in $V_1$:
\begin{equation}
  P_{\rm{ss}}(V_1) = \int dV_2\, P_{\rm{ss}}(V_1, V_2)
  \sim \exp \left( - \frac{ V'^{\, 2}_1 }{  M_{11}^{\rm{I}} \Gamma_1 } \right)
  = \exp \left( - \frac{ \tilde{C}_1 V'^{\, 2}_1 }{ 2k_B T } \right) \, ,
  \label{ss_V1}
\end{equation}
where $\hat{\bf{M}}^{\rm{I}} \equiv\hat{\bf{M}}^{-1}$, and $\displaystyle \tilde{C}_1 \equiv C_1 + \frac{ C_2 C_c }{ C_2+C_c }$ is the effective capacitance.  Based on Eq.~\ref{ss_V1}, we can develop a naive interpretation (``description (A)''), where the masked circuit is treated as a single-RC circuit with capacitance $\tilde{C}_1$ and unmodified $R_1$ ad $I_1$.  Thus $R_2$ and $I_2$ are neglected intentionally.  This effective single-RC circuit can give the correct steady-state distribution in $V_1$.  Alternatively, one can regard the time series of $V_1$ as that from a single-RC circuit, as the effective resistance and capacitance can be derived from its power spectrum, which can be shown to be identical with $R_1$ and $\tilde{C}_1$, respectively.

The total entropy change of this {\it gedanken} single-RC circuit, during an infinitesimal timestep, is
\begin{equation}
  d\tilde{S}^{\rm{(A)}}_{1\rm{tot}} = dt\, V_1 \tilde{i}_1 / T - k_B\ln \frac{ P_{\rm{ss}}(V_1(t+dt)) }{ P_{\rm{ss}}(V_1(t)) }\, ,
  \label{dS1_A}
\end{equation}
where the first term on the RHS represents a ``virtual'' dissipation, as $\tilde{i}_1 \equiv I_1 - \tilde{C}_1 \dot{V_1}$ is the virtual current going through $R_1$ in this single-RC circuit.  For the case of finite-time difference we have
\begin{equation}
  \Delta \tilde{S}^{\rm{(A)}}_{1\rm{tot},\tau} = \frac{1}{T} \int_0^{\tau} dt\, (V_1 I_1 - V_{1d} \tilde{C}_1 \dot{V_1})\, .
\end{equation}
Again one finds $\Delta \tilde{S}^{\rm{(A)}}_{1\rm{tot},\tau}$ to be Gaussian, while $\langle \Delta \tilde{S}^{\rm{(A)}}_{1\rm{tot},\tau} \rangle =  I^2_1 R_1 \, \tau/T$ is the average virtual dissipation from $R_1$.  Using time-correlation functions, it is straightforward to derive its variance:
\begin{align}
  & \langle ( \Delta \tilde{S}^{\rm{(A)}}_{1\rm{tot},\tau} )^2 \rangle - \langle \Delta \tilde{S}^{\rm{(A)}}_{1\rm{tot},\tau} \rangle^2 = \nonumber \\
  & \ \ \ \ \ \ \frac{2k_B I^2_1 R_1}{T} \bigg\{ \tau - \frac{R_1 C^2_c}{C_2+C_c} + \frac{M_{12} M_{21}}{ M^2_{22}(\lambda_1 - \lambda_2) }
    \cdot [\lambda_1 (M_{22}+\lambda_2) e^{-\tau/\lambda_1} - \lambda_2 (M_{22}+ \lambda_1)e^{-\tau/\lambda_2}]  \bigg\} \, .
  \label{variance_DS1_A}
\end{align}
Therefore, FT fails with the adoption of such dissipation function, even at the small-$\tau$ limit.  Nevertheless, from Eq.~\ref{variance_DS1_A} one finds that this deviation becomes less prominent at large $\tau$.  Moreover, one can also show that the deviation from FT diminishes in the weak-coupling regime, as the deviation in variance from $2 k_B \langle \Delta \tilde{S}^{\rm{(A)}}_{1\rm{tot},\tau} \rangle $ is proportional to $C^2_c$.

\section{Reduced description (B): trace-out approach}
\label{sec_coarse_grained}

Beside the above reduced description, one can define the effective dissipation function starting from the forward transition probability of $V_1$ over infinitesimal timesteps.  It can be derived by tracing out the degree of freedom in $V_2$:
\begin{align}
   P_F(V_1(t+dt)|V_1(t)) & = \iint dV_2(t) dV_2(t+dt) P_F(\vec{V}(t+dt)|\vec{V}(t)) P_{\rm{ss}}(\vec{V})
  / P_{\rm{ss}}(V_1) \nonumber \\
  & \sim \exp \left\{ - \frac{dt}{ 2(\vec{M^{\rm{I}}_1})^{T}
  \hat{\bf{\Gamma}} \vec{M^{\rm{I}}_1} } \left[ \dot{V_1} + \frac{
  (\vec{M^{\rm{I}}_1})^{T}   \hat{\bf{\Gamma}} \vec{M^{\rm{I}}_1} V'_1 }
  { M^{\rm{I}}_{11} \Gamma_1 } \right]^2 \right\} \nonumber \\
  & \equiv \exp [ - dt\, (\tilde{M} \dot{V_1} + V'_1)^2 / (2 \tilde{\Gamma}_1) ]
  \label{P_F_V1}
\end{align}
to the lowest nonvanishing order, where $\vec{M^{\rm{I}}_1}$ is a two-dimensional vector of elements $M^{\rm{I}}_{11}$ and $M^{\rm{I}}_{12}$.  Note that this transition probability can be compared to that of a single-RC circuit with the aforementioned effective capacitance $\displaystyle \tilde{C}_1 \equiv C_1 + \frac{ C_2 C_c }{ C_2+C_c }$,
%\begin{equation}
$ \displaystyle  \tilde{M} = \tilde{R}_1 \tilde{C}_1 = \frac{M^{\rm{I}}_{11} \Gamma_1}
  {(\vec{M^{\rm{I}}_1})^{T} \hat{\bf{\Gamma}} \vec{M^{\rm{I}}_1}} =
  (M^{\rm{I}}_{11})^{-1} / \left( 1+ \alpha \right) $,
%\end{equation}
$\tilde{R}_1 = \ R_1/(1+\alpha)$, and $\tilde{I}_1 = I_1(1+\alpha)$, where
%\begin{equation}
$ \displaystyle \alpha \equiv \frac{M_{12}M_{21}}{M_{22}^2} = \frac{R_1 C_c^2}{R_2 (C_2+C_c)^2} $.
%\end{equation}
%\begin{equation}
 % \tilde{R}_1 = R_1 \Big/ \left( 1+ \frac{M_{12}M_{21}}{M_{22}^2} \right) = R_1 \Big/ \left[ 1+ \frac{R_1 C_c^2}{R_2 (C_2+C_c)^2} \right]\, ,
%\end{equation}
And the renormalized noise amplitude parameter is
\begin{equation}
  \tilde{\Gamma}_1 = (\vec{M^{\rm{I}}_1})^{T} \hat{\bf{\Gamma}} \vec{M^{\rm{I}}_1} \tilde{M}^2 = \Gamma_1 /(1 + \alpha) ,
\end{equation}
which is smaller than $\Gamma_1$.  Since $\exp [- \tilde{C}_1 V'^{\, 2}_1 / (2k_B T) ] = \exp (- \tilde{M} V'^{\, 2}_1 / \tilde{\Gamma}_1 )$,  this effective single-RC circuit also gives the correct steady-state probability distribution in $V_1$.

In this effective single-RC circuit, the reversed transition probability is derived simply by replacing $\dot{V_1}$ with $-\dot{V_1}$ and $V'_1$ with $V_1 + V_{1d}$ in Eq.~\ref{P_F_V1}.  The net dissipation and total entropy change are
\begin{align}
  d\tilde{S}^{\rm{(B)}}_{1Q} & = k_B \ln \frac {P_F(V_1(t+dt)|V_1(t)) }{ P_R(V_1(t)|V_1(t+dt))}
    = 2k_B \, dt [ V_1(V_{1d}- \tilde{M}\dot{V_1}) / \tilde{\Gamma}_1 ]\, \ \ \text{and} \\
  d\tilde{S}^{\rm{(B)}}_{1\rm{tot}} & = d\tilde{S}^{\rm{(B)}}_{1Q} - 
    k_B\ln \frac{ P_{\rm{ss}}(V_1(t+dt)) }{ P_{\rm{ss}}(V_1(t)) } =  2 k_B \, dt V_{1d} (V_1 - \tilde{M} \dot{V}_1) / \tilde{\Gamma}_1\, ,
  \label{dS1_B}
\end{align}
respectively.  Since $d\tilde{S}^{\rm{(B)}}_{1\rm{tot}}$ is linear in $V_1$ and $\dot{V}_1$, the total entropy change is Gaussian.
One can prove that FT is satisfied for $d\tilde{S}^{\rm{(B)}}_{1\rm{tot}}$.  However, violation still occurs in finite-time processes, where
\begin{equation}
\displaystyle \langle \Delta \tilde{S}^{\rm{(B)}}_{1\rm{tot},\tau} \rangle = 2k_B \tau V^2_{1d} / \tilde{\Gamma}_1 = I^2_1 R_1 \tau (1 + \alpha )/T \, ,
\end{equation}
and
\begin{align}
  \langle ( \Delta \tilde{S}^{\rm{(B)}}_{1\rm{tot},\tau} )^2 \rangle - \langle \Delta \tilde{S}^{\rm{(B)}}_{1\rm{tot},\tau} \rangle^2  
  & = \frac{2k_B I^2_1 R_1 }{T}
  \bigg\{  (1+ \alpha )^2  (\tau - M_{11}) + R_1 \tilde{C}_1
  \nonumber \\
  & \ \ \ \ + (1+ \alpha )^2
  \frac{ (\lambda_2-M_{11}) \lambda_1 e^{-\tau/\lambda_1} -
  (\lambda_1-M_{11}) \lambda_2 e^{-\tau/\lambda_2} }
  {\lambda_2 -\lambda_1} \nonumber \\
  & \ \ \ \ - \frac{(R_1 \tilde{C}_1)^2}{\lambda_2 - \lambda_1}
  \left[ \left( \frac{\lambda_2 - M_{11}}{\lambda_1} \right)
  e^{-\tau/\lambda_1} - \left( \frac{\lambda_1- M_{11}}{\lambda_2}
  \right) e^{-\tau/\lambda_2} \right] \bigg\} \, .
\end{align}
Note that, on average, the reduced description (B) gives a larger total entropy change than description (A).


\begin{thebibliography}{99}

\bibitem{Crooks99}
G. E. Crooks, Phys. Rev. E \textbf{60}, 2721 (1999).

\bibitem{Seifert12}
U. Seifert, Rep. Prog. Phys. \textbf{75}, 126001 (2012).

%\bibitem{OurMS}
%K. H. Chiang, C. W. Chou, C. L. Lee, P. Y. Lai, and Y. F. Chen, submitted.
\end{thebibliography}
\end{document}